\documentclass[cits]{PoS}
\title{Pion polarizabilities: \\No conflict between dispersion theory and ChPT}
\ShortTitle{Pion polarizabilities}
\author{Barbara Pasquini $^a$, \speaker{Dieter Drechsel} $^b$, and Stefan Scherer $^b$\\
\llap{$^a$} Dipartimento di Fisica Nucleare e Teorica, Universit\`{a} degli Studi di Pavia,\\
and INFN, Sezione di Pavia, Italy \\
\llap{$^b$} Institut f\"{u}r Kernphysik, Universit\"{a}t Mainz, Germany\\
E-mail: \email{pasquini@pv.infn.it}, \email{drechsel@kph.uni-mainz.de}, \email{scherer@kph.uni-mainz.de}}
\abstract{Recent attempts to determine the pion polarizability
by dispersion relations yield values that disagree with the predictions
of chiral perturbation theory. These dispersion relations are based on
specific forms for the absorptive part of the Compton amplitudes.
The analytic properties of these forms are examined, and the strong
enhancement of intermediate-meson contributions is shown to be
connected to non-analytic structures.\\

PACS\{11.55.Fv,13.40.-f,13.60.Fz\}}
\FullConference{6th International Workshop on Chiral Dynamics\\
         July 6-10 2009\\
         Bern, Switzerland}
\begin{document}
\section{Introduction}
The polarizabilities of a composite system such as the pion are
elementary structure constants, just as its size and shape.
They can be studied by applying electromagnetic fields to the
system, that is, by the Compton
scattering process $\gamma + \pi \rightarrow \gamma + \pi$ or the
crossed-channel reactions $\gamma + \gamma \rightleftarrows \pi + \pi$.
The low-energy theorem expresses the zeroth- and first-order terms of the amplitude
by the charge and the mass of the pion. The second-order terms in the
photon energy describe the response of the pion's internal
structure to an external electric or magnetic dipole field,
they are proportional to the electric ($\alpha$) and magnetic
($\beta$) dipole polarizabilities. In this contribution
we concentrate on the polarizabilities determined by forward ($\alpha + \beta$) and
backward ($\alpha - \beta$) scattering.\\
Within the framework of the partially conserved axial-vector (PCAC)
hypothesis and current algebra, the
polarizabilities of the charged pion were related to the
radiative decay $\pi^+\to e^+\nu_e\gamma$~\cite{Terentev:1972ix}.
Chiral perturbation theory (ChPT) at
leading non-trivial order, ${\cal O}(p^4)$,
confirmed this result, $\alpha_{\pi^+}=-\beta_{\pi^+}\sim \bar l_\Delta$~\cite{Bijnens:1987dc},
where $\bar l_\Delta\equiv(\bar l_6-\bar l_5)$ is a linear
combination of scale-independent parameters of the Gasser and
Leutwyler Lagrangian~\cite{Gasser:1983yg}. At ${\cal O}(p^4)$
this combination is related to the ratio of the
pion axial-vector form factor $F_A$ and the vector form factor
$F_V$ of radiative pion beta decay, $F_A/F_V={\bar{l}}_\Delta/6$ ~\cite{Gasser:1983yg}.
Once this ratio is known, chiral symmetry makes an {\emph {absolute}}
prediction at ${\cal O}(p^4)$, $\alpha_{\pi^+}=2.64\pm 0.09$, here and
in the following in units of $10^{-4}\, \mbox{fm}^3$.
Corrections to this leading-order result were
calculated at ${\cal O}(p^6)$ and turned out to be rather small~\cite{Burgi:1996qi,Gasser:2006qa}.
This makes the following predictions for the polarizabilities a very significant
test of ChPT~\cite{Gasser:2006qa}:
\begin{eqnarray}
\alpha_{\pi^+} + \beta_{\pi^+} &=& 0.16 \pm 0.1\,,
\label{eq:1.1}\\
\alpha_{\pi^+} - \beta_{\pi^+} &=& 5.7 \pm 1.0\,.
\label{eq:1.2}
\end{eqnarray}
The results of ChPT are in sharp contrast with recent predictions based on
dispersion relations (DRs)~\cite{Fil'kov:2005ss},
\begin{equation}
\alpha_{\pi^+} - \beta_{\pi^+}= 13.60 \pm 2.15 \,.
\label{eq:1.3}
\end{equation}
In this work, the dispersion integrals are saturated by various meson contributions in the
$s$ and $t$ channels. The free parameters are fixed by the
masses, total widths, and partial decay widths of these mesons at
resonance. However, the extrapolation to energies below and above resonance
is performed with specific resonance shapes whose analytic properties
leave room for considerable model dependence.\\
In the present contribution we address the conflicting results obtained by ChPT and DRs.
Section~\ref{sec:Mandelstam} gives a brief introduction to the Mandelstam
plane and the kinematics used to describe the scattering amplitudes. In
Sec.~\ref{sec:Model} we present the elements of the DRs
used by Fil'kov and Kashevarov~\cite{Fil'kov:2005ss}
and investigate the analytic structure of this model along the lines of our
earlier work~\cite{Pasquini:2008ep}.
Our conclusion in Sec.~\ref{sec:Conclusion} summarizes the arguments against
the predictions of Ref.~\cite{Fil'kov:2005ss}.
\section{Kinematics and Mandelstam Plane}
\label{sec:Mandelstam}
\begin{figure}
\begin{center}
\includegraphics[width=0.49\textwidth]{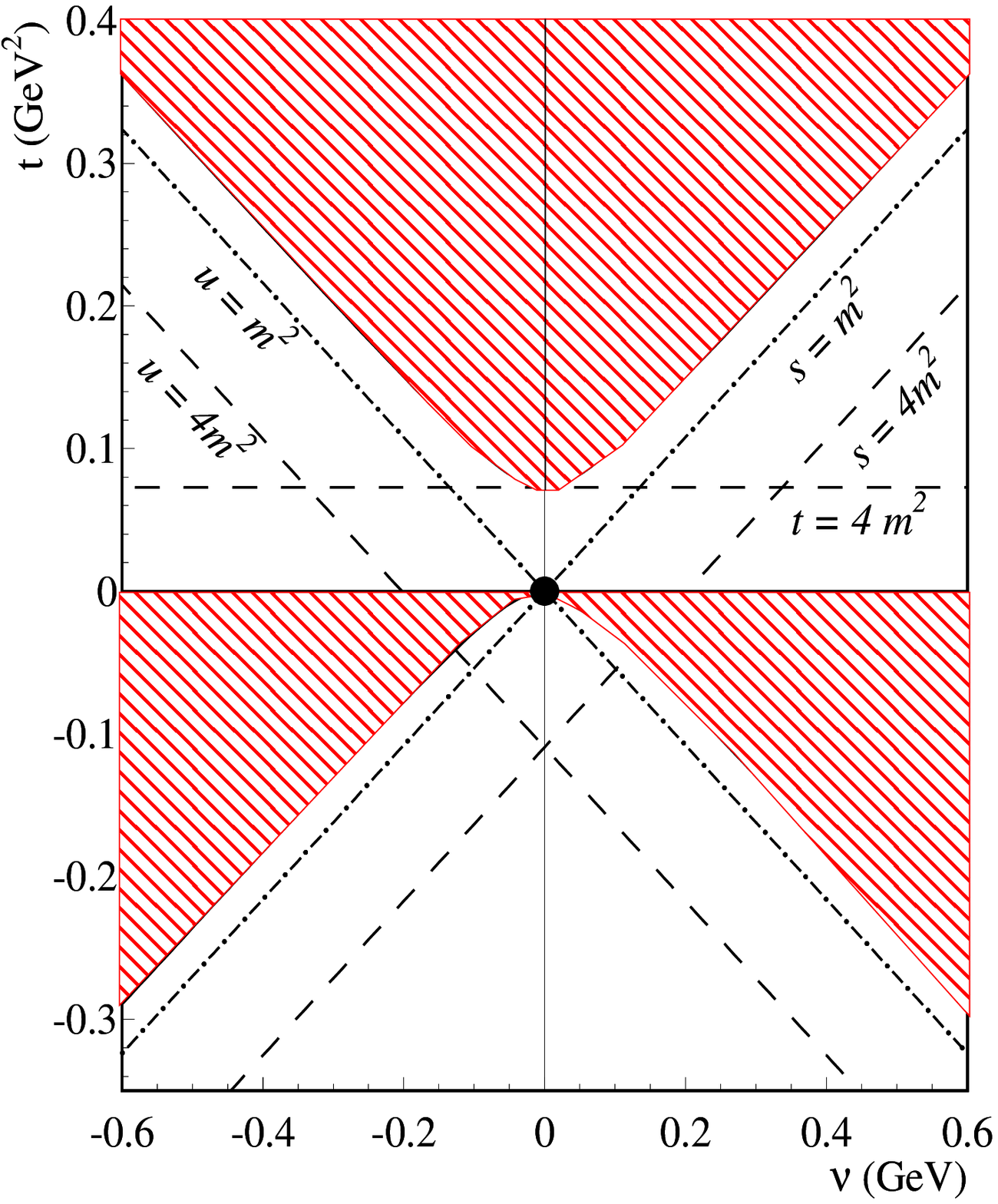}
\includegraphics[width=0.49\textwidth]{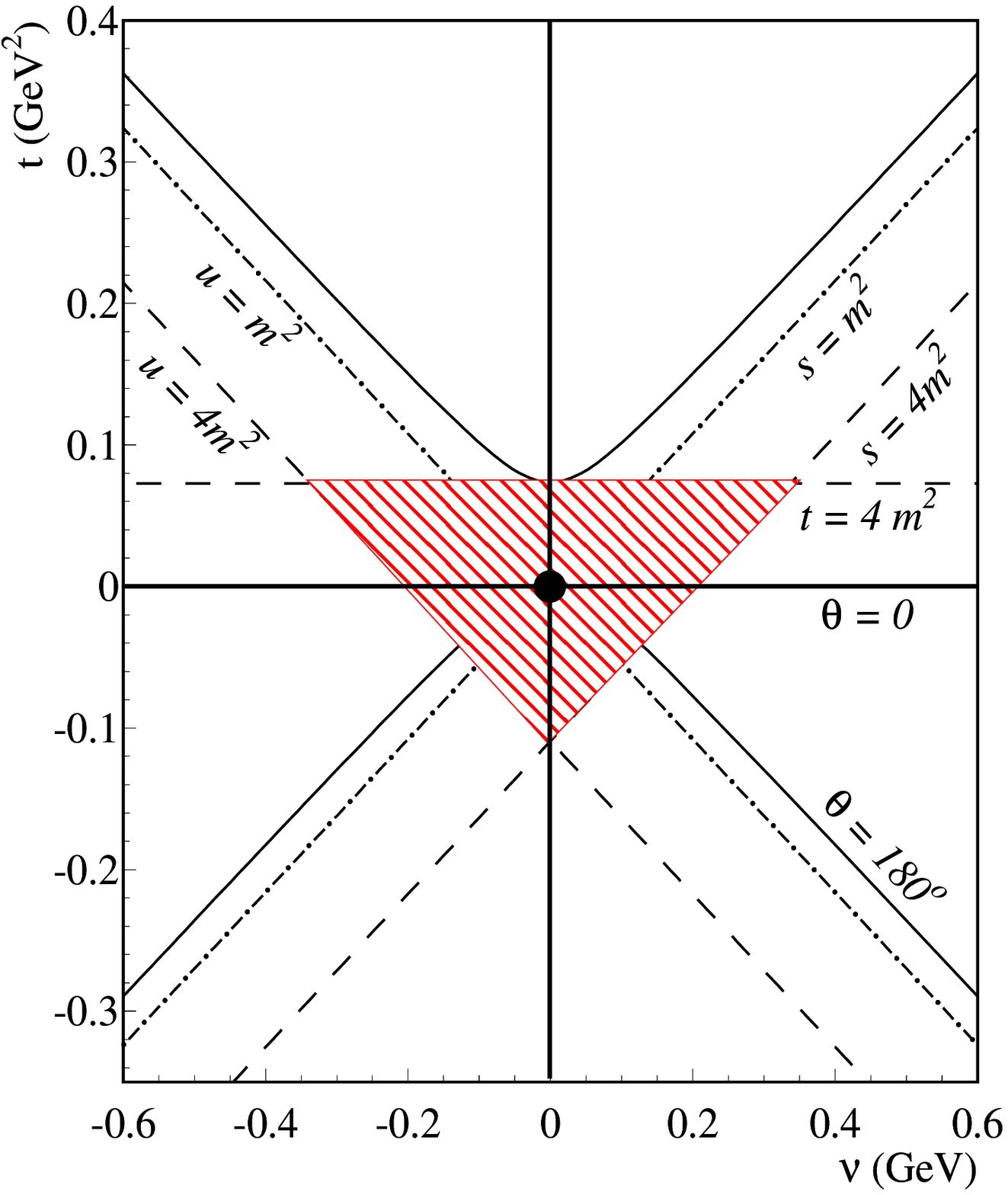}
\end{center}
\caption{The Mandelstam plane for Compton scattering on the pion.
Left panel: The hatched areas for $s>m^2$, $t>4m^2$, and $u>m^2$ show the physical regions
in the $s$, $t$, and $u$ channels, respectively.
Right panel: The hatched area forms the triangle bounded by the thresholds of two-pion
production, $s=4m^2$, $t=4m^2$, and $u=4m^2$.
See text for further explanation.}
\label{fig:mandelstam}
\end{figure}

The Mandelstam variables for Compton scattering,
$\gamma(k) + \pi(p) \rightarrow \gamma(k') + \pi(p')$,
\begin{equation}
s=(k+p)^2\, ,\quad t=(k-k')^2\, ,\quad u=(k-p')^2 \, , \label{eq:2.1}
\end{equation}
are constrained by $s+t+u=2m^2$, where $m$ is the pion mass.
The crossing-symmetric variable $\nu$ is defined by
\begin{equation}
\nu=\frac{s-u}{4m}\, . \label{eq:2.2}
\end{equation}
The two Lorentz-invariant variables $\nu$ and $t$ span the Mandelstam
plane shown in Fig.~\ref{fig:mandelstam}. They are related to the
initial ($E_\gamma$) and final ($E'_\gamma$) photon lab energies and
to the lab scattering angle $\theta$  by
\begin{eqnarray}
\nu&=&E_\gamma+\frac{t}{4m}=\frac{1}{2}(E_\gamma+E'_\gamma),\nonumber\\
t&=&-4E_\gamma \, E'_\gamma \, \sin^2 (\theta /2)
= -2m(E_\gamma-E'_\gamma). \label{eq:2.3}
\end{eqnarray}
The scattering matrix can be expressed by 2 independent amplitudes,
denoted by $M^{++}(\nu,t)$ and $M^{+-}(\nu,t)$, the superscripts indicating
the helicity of the photons in the channel
$\gamma + \gamma \rightarrow \pi + \pi$. Because of the
crossing symmetry they satisfy the relation
$M^{+ \pm}(-\nu,t)=M^{+ \pm}(\nu,t)$.\\
The Mandelstam plane is displayed in Fig.~\ref{fig:mandelstam}. The hatched
areas in the left panel show the physical regions in the $s$, $t$, and $u$ channels. The
dashed lines mark the onset of inelasticities due to the thresholds of
two-pion states with invariant mass $\geq 4m^2$ in the respective
channels. The polarizabilities are obtained in the
quasi-static limit, that is,
in the limit of approaching the origin of the Mandelstam plane,
$s=u=m^2$ or $\nu=t=0$ (solid circle).
The polarizabilities may be calculated by dispersion integrals
running along lines through the origin, e.g., $\nu=0$, $t=0$, and $u=m^2$.
The right panel of Fig.~\ref{fig:mandelstam} highlights the triangle
below the thresholds: within the dashed area,
the non-Born scattering amplitudes for real $t$ and $\nu$ are real functions
without any singularities. Therefore, these amplitudes can
be Taylor expanded about the origin of the Mandelstam plane,
with a convergence radius given by the border lines of the triangle.
The leading terms of these Taylor series describe Thomson scattering,
the polarizabilities can be read off the subleading terms.
Contrary to this basic requirement, the model of Ref.~\cite{Fil'kov:2005ss}
contains lines of square-root singularities passing through the internal
triangle. The vector mesons are modeled with a factor $1/\sqrt{s}$, which
also leads to a factor $1/\sqrt{u}$ by the crossing symmetry, and the
scalar $\sigma$ meson contains a factor $1/\sqrt{t}$ leading to a divergence
at the origin of the Mandelstam plane, at which point the
polarizabilities are determined. In the following we concentrate
on the vector mesons, in particular the $\rho (770)$. The case of the
$\sigma$ meson has been discussed in Ref.~\cite{Pasquini:2008ep}, most of the other intermediate
meson states in the work of Ref.~\cite{Fil'kov:2005ss} have similar singularities.
\section{Model of Fil'kov and Kashevarov for Vector Mesons}
\label{sec:Model}
The model of Ref.~\cite{Fil'kov:2005ss} describes the contribution of a vector meson
$V=\{\rho, \omega\}$ to Compton scattering by an energy-dependent
coupling constant $g(s)$ at the vertex $\gamma + \pi \leftrightarrows V$
and a vector meson propagator $1/\{(M-i\Gamma (s)/2)^2-s\}$, with $M$ the mass of
the vector meson and $\Gamma (s)$ its energy-dependent width. The term
quadratic in $\Gamma (s)$ is neglected (small-width approximation).
In the notation of Ref.~\cite{Fil'kov:2005ss}, the vector meson
contributions to the amplitudes take the form
\begin{equation}
\label{eq:3.1}
M^{+-}(s)=\frac{4\, g(s)^2}{M^2-s-iM\Gamma (s)}\,,
\quad  M^{++}(s)= -s \, M^{+-}(s)\,.
\end{equation}
The width $\Gamma (s)$ has the energy dependence of a P wave at threshold,
\begin{equation}
\label{eq:3.2}
\Gamma(s) = \bigg(\frac {s-4 \, m^2}{M^2-4 \, m^2}\bigg)^{3/2}\,
\Gamma_0 \,,
\end{equation}
with $\Gamma_0$ the width of the vector meson $V$ at resonance,
$s=M\,^2$. Furthermore, the square of the energy-dependent coupling
constant, $g(s)^2$, is defined with an $s^{-1/2}$ singularity in order
to obtain convergence of the dispersion integrals at large energies,
\begin{equation}
\label{eq:3.3}
g(s)^2 = \frac {6 \pi M} {s^{1/2}} \bigg ( \frac {M}{M^2-m^2}
\bigg)^3\,\Gamma_{\gamma}\,,
\end{equation}
with $\Gamma_{\gamma}$ the partial decay width for $V\rightarrow \pi + \gamma$.\\
The vector meson contributions to the polarizabilities are derived
from the amplitudes by
\begin{equation}
\label{eq:3.4}
\alpha + \beta=\frac {m} {2 \pi} M^{+-}(s=m^2),\quad
\alpha - \beta= \frac {1} {2 \pi m} M^{++}(s=m^2)\,.
\end{equation}
Combining these equations with Eq.~(\ref{eq:3.1}), we obtain
\begin{equation}
\label{eq:3.5}
\alpha+\beta= -(\alpha- \beta) \quad \Rightarrow \quad \alpha= 0\,.
\end{equation}
The (quasi-static) electromagnetic transition from the pion ($J^P=0^-$)
to the intermediate vector meson ($J^P=1^-$) is a magnetic
dipole transition yielding a paramagnetic contribution to $\beta$ and, as a consequence,
a ratio $R=(\alpha- \beta)/(\alpha+\beta)=-1$. To the contrary,
Fil'kov and Kashevarov predict a ratio $R\approx-20$ and a large electric
polarizability $\alpha$ (see Table~\ref{tab:1}). Even more surprising,
the latter carries a negative sign, a result
possible only in a relativistic quantum field theory. In conclusion, the results of Ref.~\cite{Fil'kov:2005ss} are
at variance with Eq.~(\ref{eq:3.5}).\\
\begin{table}
\begin{center}
\begin{tabular}{|l|cc|cc|}
\hline
meson & $\alpha+\beta$ & $ \alpha - \beta$ & $\alpha$ & $\beta$ \\
\hline\hline
$\rho $ & $0.063$ & $-1.15$  & $ -0.54$ & $0.61$ \\
$\omega$ & $0.721$ & $-12.56$ & $- 5.92$ & $6.64$ \\
\hline
\end{tabular}
\end{center}
\caption{\label{tab:1}Results of Ref.~\cite{Fil'kov:2005ss}
for the contributions of $\rho$ and $\omega$ mesons to the
polarizabilities of charged and neutral pions, respectively, in units of $10^{-4}\, {\rm {fm}}^3$.}
\end{table}
\begin{figure}
\begin{center}
\includegraphics[width=0.49\textwidth]{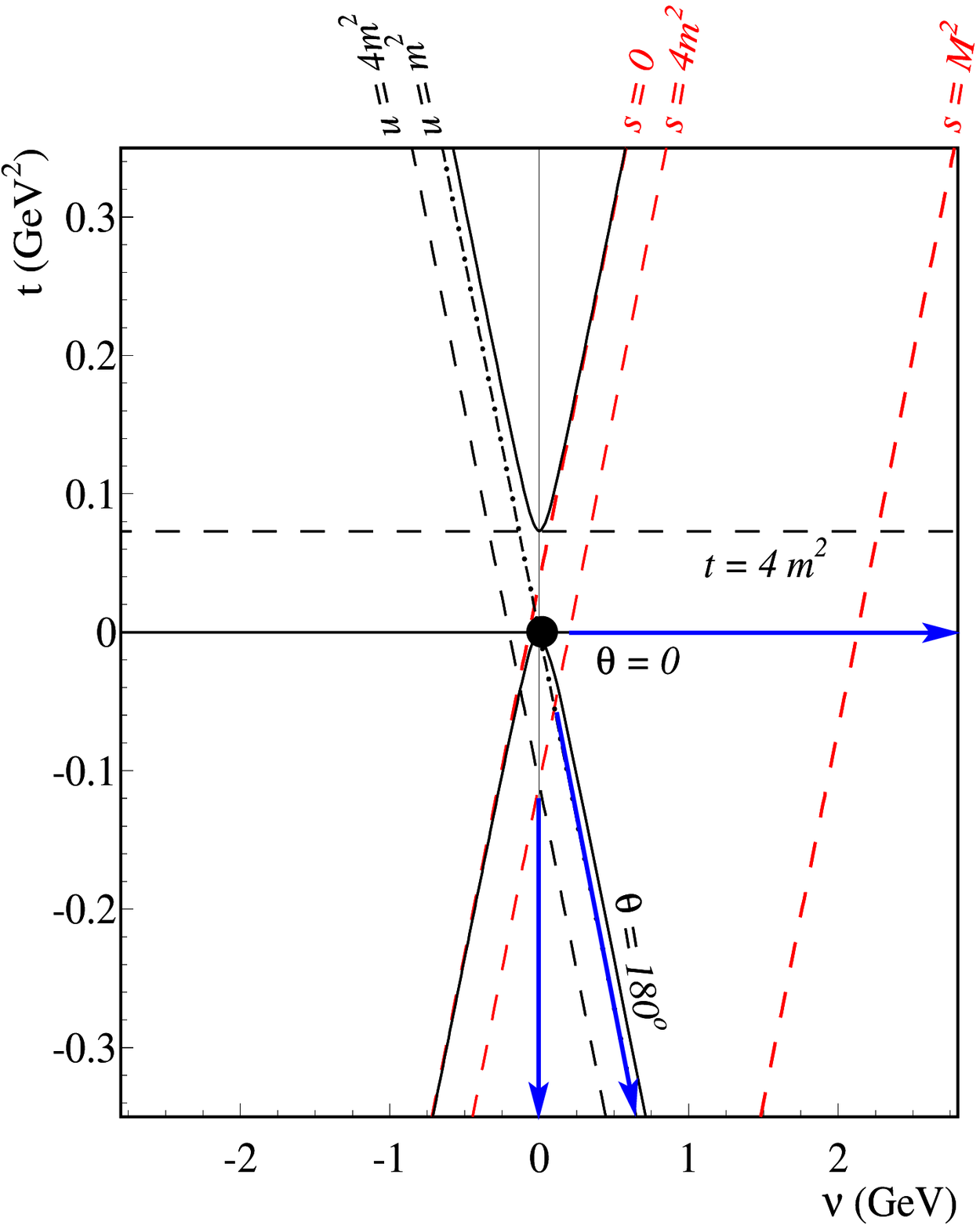}
\includegraphics[width=0.49\textwidth]{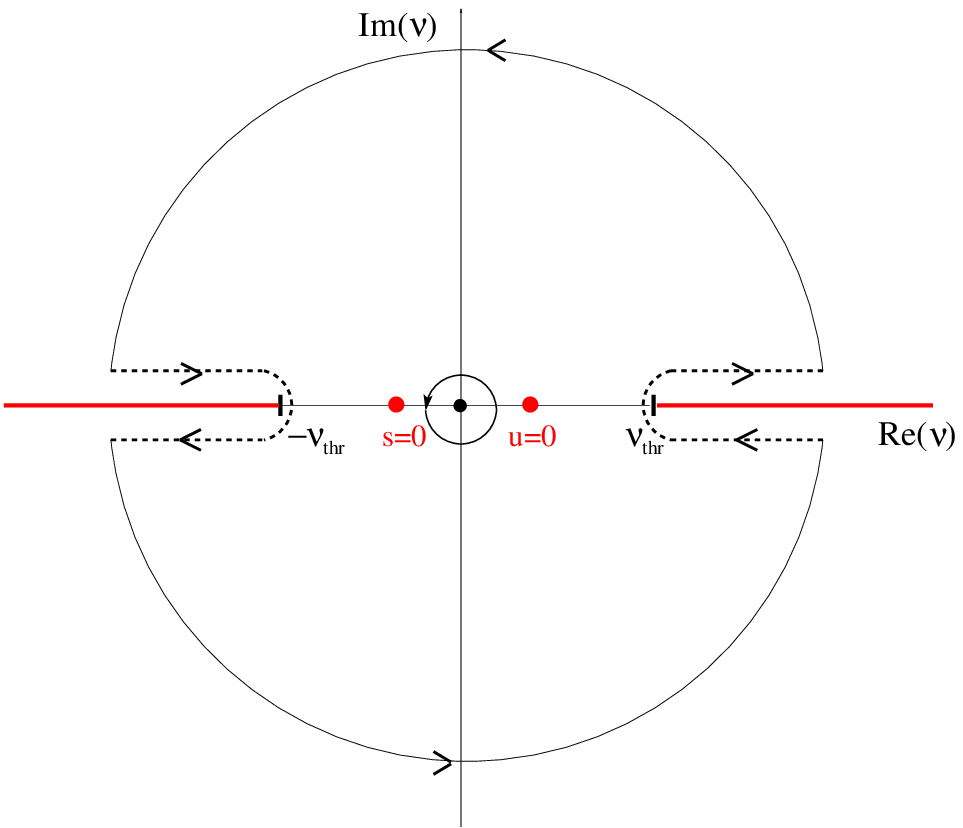}
\end{center}
\caption{Dispersion relations for Compton scattering on the pion.
Left panel: The $s$ channel singularities introduced by Ref.~\cite{Fil'kov:2005ss} are the vector meson
resonance near $s=M^2$, the onset of the physical cut at $s=4m^2$,
and the onset of the unphysical cut at $s=0$. In order to determine the
polarizability at $(\nu=0, t=0)$ (solid circle), the dispersion integral
can be obtained along different paths, e.g., $t=0$, $u=m^2$, and $\nu=0$.
Right panel: Contour integrals in the complex $\nu$ plane, related to forward DR (path along $t=0$).
On provision that the physical Riemann sheet is free of singularities except for
the physical cuts describing particle production, the Cauchy integral along the
small circle about $\nu=0$ can be replaced by the dispersion integrals
over the discontinuities along the cuts (dashed lines) and a big circle
whose contribution should vanish for an infinitely large radius. However, the model of Ref.~\cite{Fil'kov:2005ss}
contains additional square-root singularities at $s=0$ and $u=0$.
See text for further explanation.}
\label{fig:dispersion}
\end{figure}
The discussed discrepancy is related to the unphysical singularities
at $s=0$, $t=0$, and $u=0$, very close to the
origin of the Mandelstam plane at which the polarizabilities are
determined. The left panel of Fig.~\ref{fig:dispersion} displays the singularities
of the model in the $s$ channel, the unphysical singularity at $s=0$,
the threshold for two-pion production at $s=4m^2$, and the vector-meson
pole position near $s=M^2$. Also shown are the paths of integration for
three DRs, along $t=0$, $u=m^2$, and $\nu=0$. Fil'kov and Kashevarov evaluate
the dispersion integral over the physical cut, from $s=4m^2$ to infinity,
they ignore the $s^{-1/2}$ singularity and the unphysical cut resulting from it.
However, the factors $s^{-1/2}$ and $u^{-1/2}$ are not small details
but necessary to fulfill the premises of Titchmarsh's theorem, the square
integrability of the amplitude along any line parallel to the real axis.
Moreover, even for a model fitting the data in the physical region, one should not
trust the extrapolation to the origin of the Mandelstam plane if the model has
a singularity at or close to this point. Setting the imaginary parts of the amplitudes equal to zero for
Re$[s]<4m^2$,  Fil'kov and Kashevarov
tacitly annul Titchmarsh's theorem and at the same time
create a non-analytic amplitude, for which no DRs exist. The right panel of Fig.~\ref{fig:dispersion}
illustrates the problem in the complex $\nu$-plane. Let us start
from a contour integral along a small circle about $\nu=0$. If the function $F(\nu)$ is analytic within the
circle, the function at $\nu=0$ is given by Cauchy's integral,
\begin{equation}
\label{eq:3.6}
F(0)=\frac {1}{2 \pi i} \oint \frac {F(\nu')} {\nu'}\, d \nu'\,.
\end{equation}
Assuming that the physical sheet contains only the singularities of two-pion production
at $\nu=\pm \nu_{\rm {thr}}=\pm \frac {3}{2} \,m$, we can blow up the small circle to
the big contour in the right panel of Fig.~\ref{fig:dispersion}, without changing the value of the
integral, $F(0)$. On condition that the function is square integrable, the
contribution of the large circle (solid part of the big contour) vanishes if its radius
goes to infinity. The function $F(0)$ is then determined
by the integrals over the discontinuity of Im[$F(\nu)$] on the physical
cuts $-\infty < {\rm {Re}}(\nu) \leq -\nu_{\rm{thr}}$ and
$\nu_{{\rm{thr}}} \leq {\rm {Re}}(\nu) <\infty$
(dashed part of the big contour). Because the discussed model
has additional square-root singularities at
$\nu=\pm \frac {1}{2} \,m$, additional {\emph {unphysical}} cuts appear, which
we suggest to draw along the real axis such that at least the neighborhood of
$\nu=0$ remains an analytic region. With these provisions Titchmarsh's
theorem assures that DRs exist, however at the expense of introducing
unphysical cuts. \\
Although the square-root factors are essential to provide the required
convergence at infinity, Fil'kov and Kashevarov neglect the large contributions
from the associate discontinuities. As an example, such neglect could be
turned into the statement Im$[F(\nu)]=0$ if
$-\nu_{\rm{thr}}<{\rm{Re}}[\nu]<\nu_{\rm{thr}}$. The result would be two ``walls''
in the complex plane with a non-analytic strip in between. Quite generally,
any statement that the imaginary part of an analytic function be zero in
a certain area makes this area (and therefore the function itself) a
non-analytic one, because an analytic function can only contain
{\emph {isolated}} singular points.\\
\begin{figure}
\begin{center}
\includegraphics[width=0.49\textwidth]{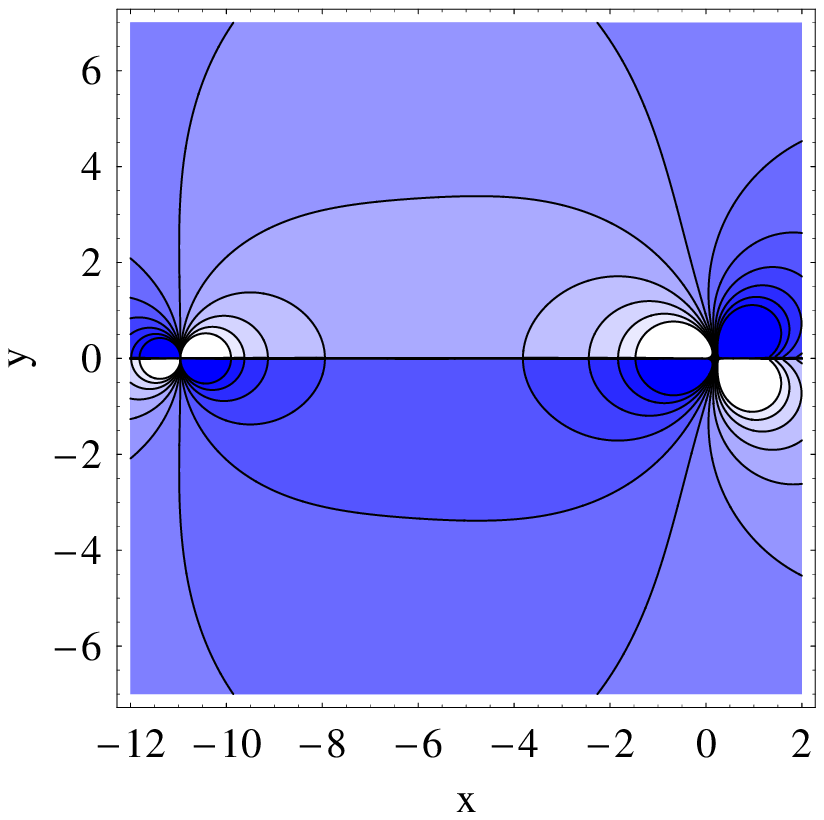}
\includegraphics[width=0.49\textwidth]{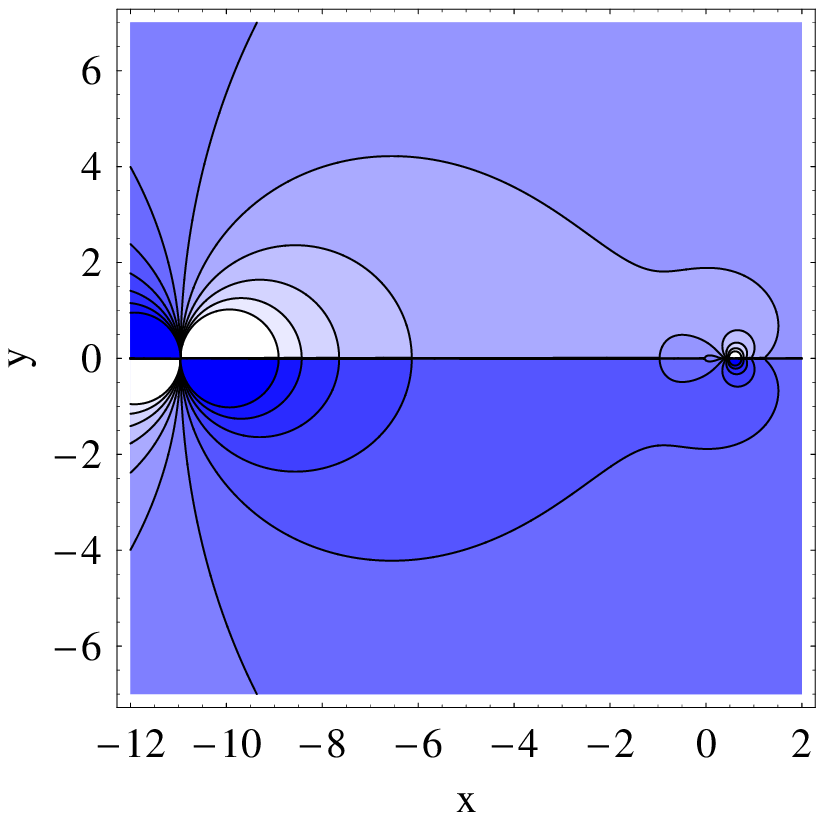}
\end{center}
\caption{Contour plots for the imaginary parts of the $\rho$ meson contributions to
$M^{+-}(s)$ (left panel) and $M^{++}(s)$ (right panel) in the complex plane
with coordinates $s=x+iy$, all in units of ${\rm {GeV}}\,^2$.
}
\label{fig:Im}
\end{figure}
The analytic structure of the model amplitudes is further illustrated in Fig.~\ref{fig:Im}.
The left panel shows the contour plot of Im$[M^{+-}(s)]$ in the complex plane, $s=x+iy$.
The physical (right-hand) cut with a maximum discontinuity near
$x=M^2 \approx 0.55{\rm {GeV}}\,^2$ is clearly seen. However, the
unphysical (left-hand) cut ranging from $s=0$ to the left turns out
to be of similar importance. Near $x=-11~{\rm {GeV}}\,^2$,
we also observe a ``bound state'' embedded on the left cut.
This unphysical pole is a consequence of the energy-dependent width in the
resonance propagator of Eq.~(\ref{eq:3.1}).
The forward polarizability $\alpha + \beta$ is determined at the point
$(x=m^2,y=0)$, squeezed in between the two cuts. The right panel of
Fig.~\ref{fig:Im} shows the same plot for Im$[M^{++}(s)]$, which
is related to the backward polarizability $ \alpha - \beta$. Because of the additional
factor $s$ in this amplitude, the $\rho$-meson resonance appears suppressed.\\
\begin{figure}
\begin{center}
\includegraphics[width=0.49\textwidth]{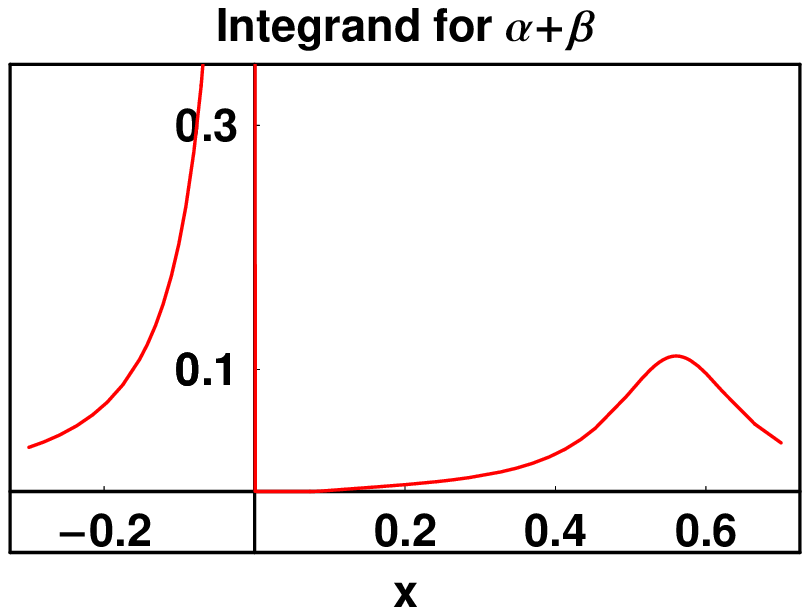}
\includegraphics[width=0.49\textwidth]{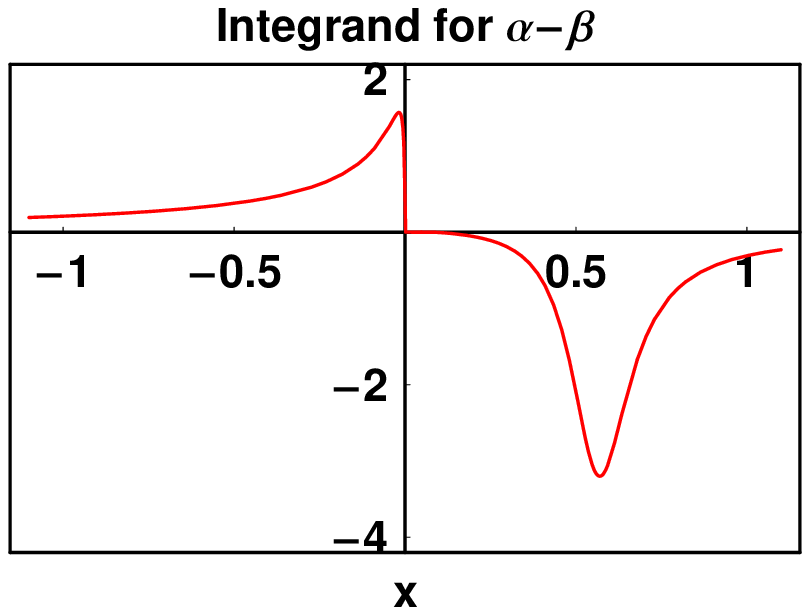}
\end{center}
\caption{The integrands of the dispersion integrals for the
polarizabilities $\alpha+\beta$ (left panel) and
$\alpha-\beta$ (right panel), as function of $x$ in units of ${\rm {GeV}}\,^2$.
See text for further explanation.}
\label{fig:Integrand}
\end{figure}
Figure~\ref{fig:Integrand} displays the integrands of the
dispersion integrals for the polarizabilities $ \alpha \pm \beta$
as function of $x={\rm {Re}}\,[s]$. The integration yields the following $s$-channel contributions
for the $\rho$ meson:
\begin{eqnarray}
\alpha+\beta &=& 0.17~({\rm {l.c.}}) + 0.03~({\rm {r.c.}})= +0.20\,,\nonumber\\
\alpha-\beta &=& 0.98~({\rm {l.c.}}) - 1.18~({\rm {r.c.}}) = -0.20\,,
\label{eq:3.7}
\end{eqnarray}
where l.c. and r.c. denote the contributions of the left and right cuts,
respectively. In agreement with Eq.~(\ref{eq:3.5}), the total result
yields a paramagnetic contribution of the vector meson to the polarizability.
In particular,  $R=(\alpha-\beta)/(\alpha+\beta)=-1$,
whereas the right-hand contributions lead to a ratio close to $R=-40$.
We note that the values for $\alpha+\beta$ given in Eq.~(\ref{eq:3.7})
include only the $s$-channel contribution of the $\rho$ meson, whereas the predictions of
Ref.~\cite{Fil'kov:2005ss} listed in Table~\ref{tab:1} contain both $s$ and 
$u$ channels.\\
Unfortunately, also somewhat modified resonance models fail to give
reasonable predictions based on dispersion relations. In Ref.~\cite{Pasquini:2008ep},
we have studied 6 such models: Model A with a constant width
$\Gamma(s)\equiv \Gamma_0$, model B with a pole at $M-i\Gamma(s)/2$,
and model C derived from model B by dropping terms quadratic in $\Gamma(s)$
as well as models A0, B0, and C0 related to the previous models by
introducing an energy-independent coupling, $g(s)\equiv g(M^{\,2})$.
As shown in Table~\ref{tab:2}, all these models contain contributions from unphysical cuts,
poles, or the ``big circle'' at infinity~\cite{Pasquini:2008ep}.
%
\begin{table}
\begin{center}
\begin{tabular}{|c||c|c|c|c||c|c|c|c|}
\hline
&\multicolumn{4}{|c||}{$\alpha+\beta$}& \multicolumn{4}{|c|}{$\alpha-\beta$}\\
\hline\hline
model & real & r.c. & l.c. & rest& real & r.c. & l.c. & rest\\
\hline
$A0$ & $0.04$ & $0.04$ & $0.00$& $-$ & $-0.04$ & $-1.04$ & $-0.08$     & $1.08$ \\
$B0$ & $0.04$ & $0.03$ & $-$     & $0.01$ & $-0.04$ & $-1.15$ & $-$     & $1.11$ \\
$C0$ & $0.04$ & $0.03$ & $-$     & $0.00$ & $-0.04$ & $-1.93$ & $-$     & $1.89$ \\
$A$   & $0.20$ & $0.05$ & $0.15$ & $-$ & $-0.20$ & $-1.06$ & $0.86$  & $-$ \\
$B$   & $0.20$ & $0.03$ & $0.17$ & $0.00$  & $-0.20$ & $-1.02$ & $0.81$ & $0.01$ \\
$C$   & $0.20$  & $0.03$  & $0.17$  & $-$ & $-0.20$ & $-1.18$& $0.97$ & $-$ \\
\hline
\end{tabular}
\end{center}
\caption{\label{tab:2}The $s$-channel $\rho$-meson contributions to the polarizabilities $ \alpha + \beta$  and
$ \alpha - \beta$  for various models described in the text.
The column ``real'' lists the values obtained directly from the real parts of the model
amplitudes, the columns ``r.c.'' and ``l.c.'' show
the contributions of the dispersion integrals over the
right and left cuts, respectively, the column ``rest'' gives contributions
from Cauchy integrals around unphysical poles and at infinity.
Within rounding errors, the three integral contributions add up to the
entry ``real''.}
\end{table}
\section{Conclusions}
\label{sec:Conclusion}
The polarizabilities of the pion are elementary structure constants and therefore fundamental benchmarks for our understanding of QCD in the confinement region. These polarizabilities have been calculated in ChPT to the two-loop order with an estimated error of less than 20~\%~\cite{Gasser:2006qa}. It is therefore disturbing that these predictions are at variance with the results of L.V.~Fil'kov and V.L.~Kashevarov by 2 standard deviations. The latter results are based on meson exchange models and performed in the framework of dispersion relations. As an example, this procedure leads to $\alpha_{\pi^+} - \beta_{\pi^+}=13.60 \pm 2.15$~\cite{Fil'kov:2005ss}, whereas ChPT predicts $5.7 \pm 1.0$~\cite{Gasser:2006qa}, all in units of $10^{-4}~{\rm {fm}}^3$. The discrepancy originates from huge contributions of intermediate meson states in the approach of Ref.~\cite{Fil'kov:2005ss}. In ChPT, on the other hand, the vector mesons enter only at ${\cal {O}}(p^6)$ through vector-meson saturation of low-energy constants. They are usually treated in the zero-width approximation and estimated to yield a much smaller effect, e.g., an $\omega$ contribution of about 1 unit to the neutral pion polarizability~\cite{Donoghue:1993kw,Bellucci:1994eb}. Moreover, Fil'kov and Kashevarov predict large contributions of vector mesons to both the electric and the magnetic polarizabilities of the pion, at variance with the fact that the transition from the pion ($J^P=0^-$) to the intermediate vector meson ($J^P=1^-$) is driven by magnetic dipole radiation leading to a (para)magnetic contribution only.
\newline
\noindent 
The apparent discrepancy between the two approaches can be traced to the specific forms for the imaginary part of the Compton amplitudes~\cite{Fil'kov:2005ss}, which serve as input to determine the polarizabilities at the Compton threshold ($s=m^2,\, t=0$) by dispersion integrals. In order to obtain amplitudes with good properties at high energies, Fil'kov and Kashevarov introduce energy-dependent coupling constants with a square-root singularity (e.g., $1/\sqrt{s}$ or $1/\sqrt{t}$). The resulting amplitudes fulfill the following conditions to set up dispersion relations:
\newline
\noindent 
(I) The amplitudes are analytic on the physical Riemann sheet except for {\emph {isolated}} points on the real axis. As an example, the $s$-channel singularities of Ref.~\cite{Fil'kov:2005ss} are situated (i) at the threshold for two-pion states ($s=4m^2$), which leads to the physical cut, and (ii) at the origin of the Mandelstam plane ($s=0$), which leads to an unphysical cut.
\newline
\noindent 
(II) The amplitudes are square integrable along any line parallel to the real 
axis, albeit at the expense of the unphysical cut starting at $s=0$.
\newline
\noindent 
Based on these conditions, Titchmarsh's theorem asserts that the real and the imaginary parts of the amplitudes are Hilbert transforms, that is, they are related by dispersion relations.
\newline
\noindent
In their actual calculations, Fil'kov and Kashevarov ignore the contribution from the unphysical cut by setting the imaginary part equal to zero ``below the threshold of two-pion production''. However one implements this statement in practice, a look at the contour plot in Fig.~\ref{fig:Im} shows that the procedure will inevitably introduce a ``wall'' in the complex plane, separating a region with finite values of the imaginary part from zero values. Because all the points on this wall become non-analytic ones, the function itself is no longer analytic and therefore condition (I) is no longer fulfilled (note: condition (I) allows only for {\emph {isolated}} singularities). Furthermore, in the region of vanishing imaginary parts, the only possible analytic function is a real constant everywhere. And unless this constant vanishes, one easily finds that also condition (II) breaks down. These arguments show the inconsistency of first introducing the factor $1/\sqrt{s}$ for convergence at large energies and later ignoring its consequence, the unphysical cut, at low energies. The results of this inconsistency are: (i) the vector meson effects are grossly overestimated and (ii) the magnetic dipole transition $\gamma + \pi \leftrightarrows \rho/\omega$ yields a {\emph {large negative}} contribution to the electric polarizability. Similar problems show up for the exchange of  other mesons. In particular, the  $1/\sqrt{t}$ factor for $\sigma$ exchange in the $t$ channel leads to a diverging amplitude $M^{++}(t)$ at $t=0$, the point at which the polarizability is to be predicted. Even apart from dispersion relations, we would not recommend to fit the data in the measurable region by functions that approach infinity at or near the point to which one wants to extrapolate. In conclusion, the reported discrepancies between ChPT and dispersion theory result from  applying the latter theory to non-analytic functions.

\end{document}